\documentclass[twocolumn,nofootinbib,aps,pra,groupedaddress]{revtex4}
\usepackage{epsfig,amssymb,amsmath}

\begin{document}
 \title{Advances in Quantum Metrology}

\author{Vittorio  Giovannetti} 
\affiliation{NEST, Scuola Normale Superiore and Istituto Nanoscienze-CNR, \\ piazza dei Cavalieri 7, I-56126 Pisa, Italy}
\author{Seth Lloyd}
 \affiliation{Massachusetts Institute of Technology -- Research Lab of
  Electronics and Dept. of Mechanical Engineering\\ 77 Massachusetts Avenue, Cambridge, MA 02139, USA}
\author{ Lorenzo Maccone}
  \affiliation{Dip.~Fisica ``A.~Volta'', Univ.~of Pavia, via Bassi 6, I-27100 Pavia, Italy }

\begin{abstract}
In classical estimation theory, the central limit theorem
implies that the statistical error in a measurement outcome can
be reduced by an amount proportional to ${n}^{-1/2}$ by repeating the measures $n$
  times and then averaging.  Using quantum
  effects, such as entanglement, it is often possible to do better,
  decreasing the error by an amount proportional to $n^{-1}$.  Quantum
  metrology is the study of those quantum techniques that allow one to
  gain advantages over purely classical approaches.  In this review,
  we analyze some of the most promising recent developments in this
  research field. Specifically, we deal with the developments of the
  theory and point out some of the new experiments. Then we look at
  one of the main new trends of the field, the analysis of how the
  theory must take into account the presence of noise and experimental
  imperfections. 
  \end{abstract}

\date{\today}

\maketitle

\begin{figure*}[hbt]
\begin{center}
\epsfxsize=.98
\hsize\leavevmode\epsffile{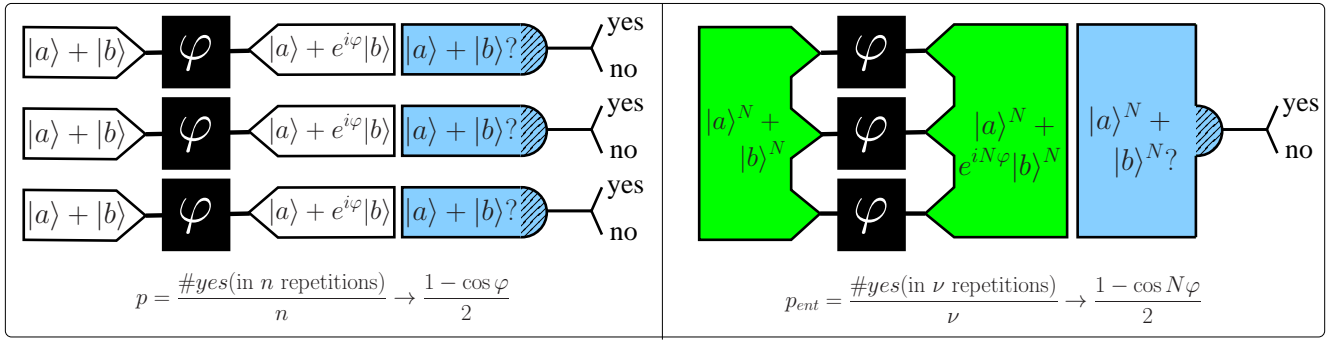}
\end{center}
\vspace{-.5cm}
\caption{Ramsey interferometry. Its aim is to  
measure an unknown relative phase $\varphi$ which is picked up by two orthogonal states $|a\rangle$, $|b\rangle$ of an atomic probing system. 
This is a generic framework
that encompasses many different interferometric
measurements (e.g. frequency-standards, magnetometry,
optical phase, etc.)  \cite{rosetta}. 
{\bf Left:}   In a conventional setup  the probe preparation consists in producing
  each atom in the superposition $|\psi_{in}\rangle=(|a\rangle+|b\rangle)/\sqrt{2}$, 
 which yields the output state  $|\psi_{\varphi} \rangle= (|a\rangle+e^{i\varphi}|b\rangle)/\sqrt{2}$ after the probing stage
 (black-boxes in the figure). The readout
consists in checking whether  $|\psi_{\varphi} \rangle$
is still in the
initial state $|\psi_{in}\rangle$.  This happens with probability $p=|\langle\psi_{in}|\psi_{\varphi}\rangle|^2=(1-\cos\varphi)/2$.
Therefore by taking the
ratio between the number of successes and the total number of
readouts, we can recover the phase $\varphi$.  
If we repeat this measurement $n$ times, the associated error on our estimation of $\varphi$, can then be evaluated by
the standard
deviation on the determination of $p$  (i.e.  $\Delta p=\sqrt{p(1-p)/n}$) and by
using error propagation
theory,  obtaining the SQL scaling
$\delta \varphi_n={\sqrt{\tfrac{p(1-p)}n}}/\big|\tfrac{\partial
    p}{\partial\varphi}\big|
=n^{-1/2}.$
{\bf Right:}   Quantum enhanced setting: A simple quantum strategy consists
in dividing the $n$ probes into groups of $N$, prepared in an
entangled state  $(|a\rangle^{\otimes N}+|b\rangle^{\otimes N})/\sqrt{2}$. 
Since each of the $N$
vectors $|b\rangle$ acquires a relative phase $\varphi$, the final
state is  $(|a\rangle^{\otimes N}+e^{iN\varphi} |b\rangle^{\otimes N})/\sqrt{2}$.
The probability that
this state is equal to the initial one is now
$p_{ent}=(1-\cos{N\varphi})/2$. Furthermore  since we have $\nu=n/N$ groups of
probes, we can repeat this procedure $\nu$ times, obtaining an error
   $\delta\varphi_{n}
   ={\sqrt{\tfrac{p_{ent}(1-p_{ent})}\nu}}/ \big|\tfrac{\partial p_{ent}}{\partial\varphi}\big|
   ={(nN)}^{-1/2}$, with a ${N}^{1/2}$ enhancement in
   precision with respect to previous case, namely, the Heisenberg
   bound for phase sensing~\cite{wineland,VBRAU92-1}.  }
\label{f:ramsey}\end{figure*}

Any measurement consists in three parts: the preparation of a probe,
its interaction with the system to be measured, and the probe readout.
This process is often plagued by statistical or systematic errors. The
source of the former can be accidental (e.g.~deriving from an
insufficient control of the probes or of the measured system) or
fundamental (e.g.~deriving from the Heisenberg uncertainty relations).
Whatever their origin, we can reduce their effect by repeating the
measurement and averaging the resulting outcomes. This is a
consequence of the central-limit theorem: given a large number $n$ of
independent measurement results (each having a standard deviation
$\Delta\sigma$), their average will converge to a Gaussian
distribution with standard deviation $\Delta\sigma/\sqrt{n}$, so that
the error scales as ${n}^{-1/2}$.  In quantum mechanics this behavior
is referred to as ``standard quantum limit'' (SQL) and is associated
with procedures which do not fully exploit the quantum nature of the
system under investigation\footnote{In quantum optics, the
  ${n}^{-1/2}$ scaling is also indicated as ``shot noise'', since it
  is connected to the discrete nature of the radiation that can be
  heard as ``shots'' in a photon counter operating in Geiger mode.}.
Notably it is possible to do better when one employs quantum effects,
such as entanglement among the probing devices employed for the
measurements, e.g. see Refs.~\cite{GIOV04,rosetta,wineland,VBRAU92-1,cavesprd,HOLL93}.  Consequently, the SQL is not a
fundamental quantum mechanical bound as it can be surpassed by using
``non-classical'' strategies.  Nonetheless, through Heisenberg-like
uncertainty relations, quantum mechanics still sets ultimate limits in
precision which are typically referred to as
``Heisenberg bounds''.  In Fig.~\ref{f:ramsey} we present a simple
example which may be useful to understand the quantum enhancement.
  Part of the
emerging field of quantum technology~\cite{OBRI09}, quantum metrology  studies these
bounds and the (quantum) strategies which allows us to attain them.
More generally it deals with measurement and discrimination procedures
that receive some kind of enhancement (in precision, efficiency,
simplicity of implementation, etc.) through the use of quantum
effects.

%%%%%%%%%%%%%%%%%%%%%%%%%%%%%%%%%%%%%%%%%%%%%%

This paper aims to review some of the most recent development of the
field.  For a more historical perspective of quantum metrology, we
refer the reader to~\cite{GIOV04}. We start by introducing some
important results on quantum estimation theory which focus on the
optimization of the probe readout.  Then we report some recent
findings obtained in the context of parameter estimation for channels,
which allows also for the optimization of the probe preparation.  It
is only at this stage that Heisenberg-like scaling is obtained.
Schemes based on filtering protocols and nonlinear effects will be also introduced.  In the
remaining of the paper we will then deal with the analysis of the
typically very fragile quantum metrology protocols in the presence of
noise.

\subsection*{Basics on Quantum   Estimation for states }\label{sec:bas}

In its simplest version a typical quantum estimation
problem~\cite{HELS67,BRAU96,BRAU94,HOLE82,HAYA05,HAYA06,PARI09}
consists in recovering the value of a continuous parameter $x$
(say the phase $\varphi$ of Fig.~\ref{f:ramsey}) which is encoded into a fixed set of states
$\rho_x$ of a quantum system $S$. 
As in the example of Fig.~\ref{f:ramsey}, we can describe it as a
two step process where we first perform a measurement on $S$, and then
extrapolate the value of $x$ with some data-processing of the
measurement results. The measurement is described by a Positive
Operator Valued Measurement (POVM) ${\cal E}^{(n)}$ of elements $\{
E_y^{(n)}\}$ (where $n$ is the number of copies of $\rho_x$ we use).
The conditional probability of getting the outcome $y$ will be then
computed as $p_n(y|x) = \mbox{Tr}[ E_y^{(n)} \rho_x^{\otimes n}]$. The
data-processing of the result $y$ will yield our estimate $z$ of the
value of $x$ and, in the most general case,  will be
characterized by assigning some conditional probabilities $p_{est}^{(n)}(z|y)$.
Ideally, we would like to have $z$
as close as possible to the parameter $x$; unfortunately,
depending on the physics of the problem and, possibly, on the selected estimation strategy,
there will often
be some residual uncertainty in the determination of the parameter.  
This uncertainty is fully characterized by the probability
$P_n(z|x) := \sum_{y} \; p_{est}^{(n)}(z|y) p_n(y|x)$,
which describes the statistical dependence of $z$ on the true value
$x$. 
It  allows us to define the 
  Root Mean Square Error (RMSE) $\delta X_n:= \sqrt{\sum_{z} [{z}-x]^2 \; P_n(z|x)}$~\cite{HELS67}  which provides a good measure of the estimation accuracy (different 
 measures are possible but will not be treated here, see e.g. Ref.~\cite{HOLE82}).
Under reasonable assumptions on the
asymptotic behavior of the estimation function (i.e.  requiring it to
be {\em asymptotically Locally Unbiased}\footnote{This means that for
  $n\to\infty$ there exists a value of the parameter $x$ for which the estimator tends to
  the correct value, and at the same time the derivative of the
  estimator at that point is unity~\cite{HELS67}. All reasonable
  estimators satisfy this condition
while the pathological ones, do not
(e.g. trivial estimation procedures that obtains the correct value of
$x$ by chance, as the case of a stopped clock 
which is correct twice per day).
  A generalization of
  Eq.~(\ref{MSE}) which applies to all estimation strategies without
  the assumption of locally unbiasedness was provided
  in~\cite{BRAU94,BRAU96}. It is obtained by replacing $\delta X_n$ on
  the lhs of Eq.~(\ref{MSE}) with its regularized version $\delta
  \tilde{X}_n=\sqrt{\sum_{z} [\tilde{z}(x)-x]^2 \; P_n(z|x)}$ in which
  $\tilde{z}(x)$ is the quantity $z$ divided by
  $\tfrac{\partial}{\partial x} \sum_{z} {z} \; P_n(z|x)$.  Even
  though the inequality for $\delta \tilde{X}_n$ refers to a larger
  number of estimation strategies, it is weaker since $\delta
  \tilde{X}_n \geqslant \delta X_n$. Note also that the regularization
  implies that $\delta \tilde{X}_n$ will diverge for the pathological
  estimators that would be excluded by the restrictions adopted in
  deriving~(\ref{MSE}).}), $\delta X_n$ can be shown to obey to the so
called Cram\'{e}r-Rao (CR) bound~\cite{CRAM46}, which implies
  \begin{eqnarray}
\delta X_n\geqslant {1}/{\sqrt{F_n(x)}}, \label{MSE}
\end{eqnarray}
where $F_n(x) := \sum_y \left(\frac{\partial p_n(y|x)}{\partial
    x}\right)^2 /p_n(y|x)$ is the Fisher information associated with
the selected POVM measurement.

Optimizing Eq.~(\ref{MSE}) with respect to all possible POVMs ${\cal
  E}^{(n)}$ one then gets the
inequality~\cite{HELS67,BRAU96,BRAU94,HOLE82,HAYA05,HAYA06,PARI09}   
 \begin{eqnarray}
   \delta X_n \geqslant  \frac{1}{\sqrt{ \max_{{\cal E}^{(n)}}
       F_n(x)}} 
   \geqslant \frac{1}{\sqrt{n \; J(\rho_x)}} \label{QCRN} \;.
\end{eqnarray}
The term on the right is the {\em quantum} Cram\'{e}r-Rao (q-CR)
bound.  It is obtained by exploiting an upper bound of $\max_{{\cal
    E}^{(n)}} F_n(x)$ in terms of the {\em quantum Fisher information}
\footnote{This is defined as $J(\rho_x):={\rm Tr}[{\cal
    R}^{-1}_{\rho_x}(\rho'_x)\rho_x{\cal R}^{-1}_{\rho_x}(\rho'_x)]$,
  where $\rho'_x=\partial\rho_x/\partial x$, and where ${\cal
    R}^{-1}_{\rho}(O):=\sum_{j,k:\lambda_j+\lambda_k\neq
    0}2O_{jk}|j\rangle\langle k|/(\lambda_j+\lambda_k)$ is the
  symmetric logarithmic derivative written in the basis that
  diagonalizes $\rho_x=\sum_j\lambda_j|j\rangle\langle j|$~\cite{HELS67}.} $J(\rho_x)$. For
instance, consider the case in which $\rho_x$ are pure states of the
form $|\psi_x \rangle = \exp[- i Hx] |0\rangle$ with $H$ an Hermitian
operator and $|0\rangle$ a reference vector.  Then, the q-CR bound
takes the simple form of an uncertainty
relation~\cite{HELS67,HOLE82,BRAU96,BRAU94},
\begin{eqnarray}
  \delta X_n \geqslant \frac{1}{2 \;\sqrt{ n} \; \Delta H }  \label{QCRN2} \;,
\end{eqnarray}
where $\Delta H := \sqrt{ \langle (H-\langle H\rangle)^2 \rangle }$ is
the spread of $H$ on $|0\rangle$.

Three things are worth stressing at this point: {\em i)} the bound in
Eq.~(\ref{QCRN}) holds for {\em all} possible POVMs, including those
which operate jointly on the $n$ copies while exploiting entanglement
resources; {\em ii)} the SQL scaling $n^{-1/2}$  on
the rhs is a direct consequence of the additivity of the quantum
Fisher information when applied to tensor states $\rho_x^{\otimes n}$,
i.e.  $J(\rho_x^{\otimes n})=nJ(\rho_x)$; {\em iii)} in the asymptotic
limit of large $n$, the q-CR bound is always achievable and the
estimation strategy which attains it can be constructed via {\em local
  measurements} and adaptive estimators, a strategy that employs only
Local Operations and Classical Communication (LOCC)
\cite{HAYA06,HAYA08,GILL99,HAYA05,FUJI06,NAGA88,NAGA89}.  This
implies that entangled resources at the measurement stage are not
necessary to achieve the q-CR bound. Local measures and some clever
classical data-processing are sufficient. It also shows that the
quantity $1/\sqrt{n J(x)}$ has a clear operational meaning and can be
used to quantify how hard the estimation problem is.

Finally, we briefly consider the multi-parameter case where $x$ is a
vector of random variables~\cite{HOLE82,HELS67,
  HAYA06,PARI09,HAYA05}.  Also in this case one can construct an
inequality in terms of the associated quantum Fisher information, but
in general it will not coincide with the ultimate achievable bound.
This inequality applies only to all $n$-body {\em separable}
measurements \cite{HAYA08} -- i.e.  POVMs which may act globally on
$\rho_x^{\otimes n}$ but whose elements can be expressed as convex
combination of tensor products of positive operators (which is a
larger class than LOCC \cite{BENN99}).  If, instead, we allow for
generic joint measurements which exploit entangled resources in a non
trivial fashion, then the bound may not hold and better performances
are expected (even though the $n^{-1/2}$
scaling still holds).

\subsection*{Quantum parameter estimation for channels}

The analysis of the previous section focused on scenarios in which the
set of states $\rho_x^{\otimes n}$ is fixed by the estimation problem.
Thus, it can only describe the last stage (probe readout) of a general
measurement. Since \eqref{QCRN} and \eqref{QCRN2} 
obey a SQL scaling $n^{-1/2}$, to obtain a quantum metrology type enhancement,
we need to  consider the other two stages  (i.e. probe preparation and
interaction). For this purpose it is useful to describe the correspondence $x \rightarrow \rho_x$
in terms of a  quantum channel $\Phi_x$ which produces $\rho_x$ when acting on the initial 
input state $\rho_0$ of the probe, through the mapping~$\Phi_x(\rho_0)=\rho_x$.
 Aim of  this section is to characterize 
 the best estimate of $x$ which can be obtained for a fixed number
$n$ of applications of $\Phi_x$,  while optimizing with respect to the
measurement, the estimation functions, {\em and} the choice of the
initial state~\cite{FUJI01,HAYA10,FUJI03,FISH01,FUJI02,BUZE99,CHIR04,DEBU05,BAGA04,mosca,GIOV06,HAYA06-1,IMAI07,BALL04,SASA02,JI08}.
This captures the basic aspects of most quantum metrology
applications~\cite{GIOV04}.

%%%%%%%%%%%%%%%%%%%%%%%%%%%%%%%
\begin{figure}[hbt]
\begin{center}
\epsfxsize=.75
\hsize\leavevmode\epsffile{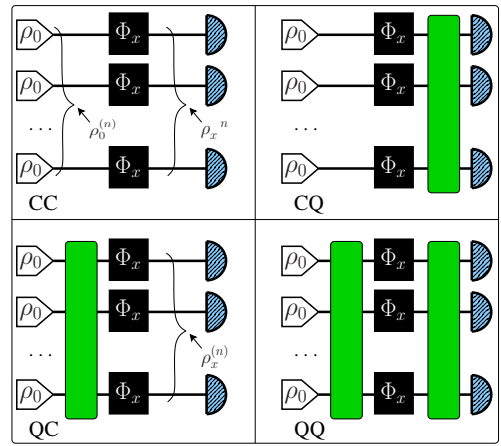}
\end{center}
\vspace{-.5cm}
\caption{Schematic representation of parallel estimation strategies. 
In the picture, a white wedge element represents an input probe entering the apparatus,
a black box represents an application of the channel $\Phi_x$, 
    a blue semicircles describes a local measurement on the probe, while a green box represents 
    an entangling operation among the probes. 
    CC (classical-input, classical-output): these are those
  strategies where the probes are prepared in a
  separable state $\rho_0^{(n)}$ (not necessarily all probes must be
  in the same state as depicted), and LOCC measurements
  at the output.  In this case, the averaging of the local results yields a decrease in the
  result's precision that at most scales as the SQL, i.e. $n^{-1/2}$. CQ (classical-input, quantum-output):
  entanglement among the probes is generated just before
  the detection. QC (quantum-input, classical-output): entanglement
  among the probes is generated before they are fed into the channel while
    no entanglement resource is employed at the detection stage which uses a LOCC
  strategy. QQ (quantum-input, quantum-output): the most general
   strategy where entanglement can be used both at the probe
  preparation and at the probe detection stages. By construction  the QQ and CC
strategies will always provide the best and the worst performances
respectively. In the case of
  estimation of unitary channels, CQ has the same $n^{-1/2}$ yield of CC, whereas
  QC and QQ can achieve the Heisenberg bound $n^{-1}$~\cite{GIOV06}. 
Non-unitary
  channels have equal or worse performances (since any non-unitary map
  can be purified into a unitary), depending on their action on the
  probes \cite{GIOV06,JI08}. 
  }
\label{f:chann-estim}\end{figure}
%%%%%%%%%%%%%%%%%%%%%%%%%%%%%%%%%

\begin{figure}[hbt]
\begin{center}
\epsfxsize=.6
\hsize\leavevmode\epsffile{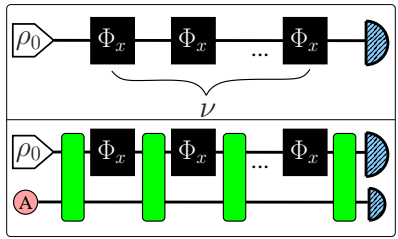}
\end{center}
\vspace{-.5cm}
\caption{{\bf Right:} Sequential (or multi-round) scheme \cite{luis,HIGG07,GIOV06,JI08,DEBU05}, where a
  single probe state samples the $n$ black boxes sequentially. {\bf
    Left:} Entanglement-assisted sequential scheme, where an external
  ancillary system A aids the estimation.These schemes have some advantages
with respect to the parallel schemes as in principle do not require to
use multipartite entanglement among $n$ different probes to achieve
sub-shot-noise scaling (furthermore they can simulate any other
schemes by properly tuning the intermediate
controls). The most general estimation
  scheme will be a combination of the sequential and of the parallel
  schemes of Fig.~\ref{f:chann-estim}. An optimization over the
  possible black-box dispositions was presented in~\cite{mosca}.  }
\label{f:seq}\end{figure}

As shown in Figs.~\ref{f:chann-estim} and~\ref{f:seq}, we can employ
different strategies depending on the choice for the initial state of
the probes, for the measurements, and for the operation we can perform
on the probed system (in parallel or sequentially)~\cite{GIOV06}.  In
full analogy to what is done in quantum channel communication theory,
we can also consider {\em entanglement assisted} schemes in which the
incoming probes are entangled with an external ancillary system $A$
which is not effected by the channel $\Phi_x$ (this is 
the most general method of performing measurements).
 Accuracy enhancements
via entanglement assisted schemes for parallel configurations have
been derived for special classes of quantum channels
$\Phi_x$~\cite{FUJI01,FISH01,BALL04,SASA02,FUJI03}.

%%%%%%%%%%%%%%%%%%%%%%%%%%%%%%%%%%%
%%%%%%%%%%%%%%%%%%%%%%%%%%%%%%%%

Optimizing with respect to all available resources can be a hard
problem. For the sake of simplicity in the following we will focus on the parallel strategies.
In this context the conventional approach is to split the optimization into
two stages~\cite{FUJI01,FISH01,FUJI03,BALL04,SASA02}. Specifically one
{\em first} fixes a (possibly entangled) input state $\rho_0^{(n)}$
and optimizes with respect to the measurements, and {\em then}
minimizes the result with respect to the $\rho_0^{(n)}$.  From the
results of the previous section it follows that the accuracy of the
first stage will not be larger than q-CR bound associated with the
selected state.  Therefore, the second optimization yields
\begin{eqnarray}
  \delta X_n \geqslant \min_{\rho_0^{(n)}}
  \frac{1}{\sqrt{J(\rho_x^{(n)})}}  \label{bound1}\;,
\end{eqnarray} 
where $J(\rho_x^{(n)})$ is the quantum Fisher information of the
output state $\rho_x^{(n)}$.
 Since the quantity
$J(\cdot)$ on the right is convex~\cite{FUJI01}, the minimum is always
achieved by choosing pure states.  In addition, due to the additivity
properties of $J(\cdot)$ it is clear that restricting the minimization
in Eq.~(\ref{bound1}) to input states which are separable with respect
to $n$ probes, one can achieve at most a $n^{-1/2}$ scaling
(SQL)~\cite{GIOV06}. Better scalings require
entangled states $\rho_x^{(n)}$. To exemplify this, we focus on the
special case in which the $\Phi_x$ perform random unitary rotations
$e^{-i x H}$.  This is a paradigmatic
model which, among others, includes the Ramsey setup of Fig.~\ref{f:ramsey}, 
and the Mach-Zehnder interferometer~\cite{rosetta,VYURK86,GIOV04,VHRAD96}, commonly used to detect
an unknown phase shift (see below).  Assume that the total number $n =
\nu N$ of probes is split in $\nu\gg 1$ groups of $N$ elements, where
each group is prepared in the same entangled input state
$|\psi_0^{(N)}\rangle$. The output state of all $n$ probes after
passing through the black-boxes is $|\psi_x^{(N)}\rangle^{\otimes \nu}:=
(e^{-iHx})^{\otimes n} |\psi_0^{(N)}\rangle^{\otimes\nu}$.  The
bound~(\ref{bound1}) can then be evaluated exploiting
Eq.~(\ref{QCRN2}) and yields~\cite{GIOV06},
\begin{eqnarray} \label{HEIS}
 \delta X_n\geqslant \frac{c}{\sqrt{\nu} N} \;,
\end{eqnarray}
where $1/c$ is the largest gap in the spectrum of the generator $H$ of
the transformation $\Phi_x$.
  For $N>1$ it presents a
$N^{1/2}$-enhancement with respect to the standard-quantum-limit.  The
term on the right is the Heisenberg bound and can be attained by
preparing the probes in each group into the entangled state
$|\psi_0^{(N)}\rangle = ( |\lambda_{max} \rangle^{\otimes N} +
|\lambda_{min} \rangle^{\otimes N} ) /\sqrt{2}$, where
$|{\lambda_{max}\rangle ,|\lambda_{min}}\rangle$ are the eigenvectors
associated to the maximum and minimum eigenvalues of $H$,
respectively\footnote{\label{fn:ram} For the Ramsey setup
  $|\psi_0^{(N)}\rangle$ corresponds to the maximally entangled state
  $(|a\rangle^{\otimes N} + |b\rangle^{\otimes N})/{\sqrt{2}}$ of
  Fig.~\ref{f:ramsey}.  In the atomic literature, where $|a\rangle$
  and $|b\rangle$ refer to different atomic levels,
  $|\psi_0^{(N)}\rangle$ often is referred to as ``spin-squeezed
  state''~\cite{ueda,cirac} or ``Schr\"{o}dinger-cat
  state"~\cite{LEIB03,LEIB04,LEIB05}. It has been experimentally
  demonstrated using trapped ions~\cite{spp,LEIB03,LEIB04,LEIB05},
  Bose-Einstein condensates~\cite{spp1}, macroscopic atomic
  ensembles~\cite{polzik}, and NMR~\cite{JONE09}.}.  From the previous
section it follows that the threshold~(\ref{HEIS}) is asymptotically
attainable for $\nu \gg 1$ by exploiting POVMs which act locally on
the blocks and a maximum likelihood approach.
Alternatively~\cite{GIOV06} 
it is also achievable via adaptive strategies based on
POVM which acts locally on each probe (so that the yield of QC
strategies coincides with the one of QQ strategies).

In contrast, the situation is more complicated for finite $\nu$,
specifically for $\nu=1$ where Eq.~(\ref{HEIS}) would yield the
impressive $n^{-1}$ scaling for the RMSE of the Heisenberg
bound.  For the special case in which $\Phi_x$
induces unitary transformations, this problem was analyzed in
Refs.~\cite{JI08,DEBU05} where a scaling of order $n^{-1} \log n$ was
achieved by means of local adaptive strategies. In Ref.~\cite{JI08} it
was also observed that if the channels $\Phi_x$ are {\em
  programmable}\footnote{\label{fn:prog} The map $\Phi_x$ is called
  programmable if one can express it in terms of a constant
  interaction with an external ancillary system $B$ whose initial
  state encodes the dependence on the parameter $x$,
  i.e.~$\Phi_x(\cdot) = \mbox{Tr}_B \{V [\sigma_x \otimes (\cdot)]
  V^\dag\}$ where $V$ is a unitary transformation acting on $B$ and
  $S$, $\sigma_x$ is the state of $B$, and $\mbox{Tr}_B\{\cdot\}$ is
  the partial trace on $B$~\cite{NIEL97}.} then a sub-shot-noise
scaling of the accuracy is not allowed.  Examples for which the no-go
theorem of~\cite{JI08} hold are provided by the class of classical
channels (i.e. quantum channels $\Phi_x$ which cannot propagate
quantum information) and by the class of depolarizing
channels~\cite{SASA02}: for these families of maps the best estimation
accuracy of the parameter $x$ will scale at most as $n^{-1/2}$.
  Recently it was also pointed
out that for finite $\nu$, the two-step optimization approach adopted
in the derivation of Eq.~(\ref{bound1}) in general fails to provide
the achievable bound~\cite{HAYA10,HAYA10-1}.  Instead one must adopt a
min-max optimization scheme, minimizing the maximum RMSE, which always
yields attainable bounds. It gives slightly worse performance than the
two-step approach, but (apart from numerical prefactors) it maintains
the same $n^{-1}$ scaling (at least when the quantum channels
$\Phi_x$ induce unitary transformations).  In concluding, we also
briefly mention that analogous $n^{-1}$ scalings for specific problems
of quantum channel estimation have been obtained by using error
measures which are substantially different from the RMSE adopted here
(e.g. from those results one would not be able to infer the exact
$n^{-1}$ scaling for the RMSE which follows from the q-CR bound).  In
particular Refs.~\cite{BUZE99,BAGA04,CHIR04,HAYA06-1,mosca} consider the
problem of estimating unitary rotations in finite dimensional systems
using a metric introduced by Holevo~\cite{HOLE82}.

\subsection*{Applications in Quantum interferometry} 

The prototypical example of a quantum interferometric application of a quantum estimation procedure is
provided by the Mach-Zehnder interferometer.  In this setting two
input optical modes are merged at a first 50-50 beam splitter,
propagate along two paths of different length accumulating an
(unknown) relative phase shift $\varphi$, and are then merged at a
second 50-50 beam splitter.  The goal is to recover the value of
$\varphi$ by measuring the signals emerging from the interferometer,
while employing a limited amount of resources (i.e.  by setting an
upper limit $N$ on either the maximum number or on the average number
of the photons entering the interferometer at each experimental run).
Indicating with $a$ and $b$ the annihilation operators associated with
the two internal paths of the interferometer, the problem of
recovering $\varphi$ reduces to estimating a channel
$\Phi_{x=\varphi}$ which induces a unitary rotation $e^{- i x H}$ with
$H = (a^\dag a - b^\dag b)/2$ being the effective system Hamiltonian,
e.g. see Refs.~\cite{VYURK86,rosetta,VHRAD05}.  With this
identification the two-step optimization strategy that brought us to
Eq.~(\ref{HEIS}) can be used to set a lower bound on the RMSE.  First
consider the situation in which generic POVM measurements are
performed on $\nu$ independent preparations of the interferometer.  In
this case Eq.~(\ref{QCRN2}) yields the following bound
\begin{eqnarray}
  \delta \varphi_\nu \geqslant \min_{|\Psi \rangle} 
  \frac{1}{2 \;\sqrt{\nu} \; \Delta H }  \label{QCRN33} \;,
\end{eqnarray}
where the minimization is performed over the set of input states
$|\Psi\rangle$ which satisfy the selected photon number constraint
(i.e. the maximum number, or the average number constraint), and where
$\Delta H$ is the associated energy spread.  Under both constraints,
the optimal input $|\Psi\rangle$ is provided by a state which at the
level of the internal modes of the interferometer, can be expressed as
a ``NOON" state \cite{bollinger,rosetta,BOTO01}, a superposition of
the form $(|N,0\rangle+|0,N\rangle)/\sqrt{2}$, in which $N$ photons
are propagating along the first or the second optical path, e.g.~see
Refs.~\cite{DURK07}.  NOON states are the formal analogue of the
spin-squeezed states that achieve the Heisenberg bound in a Ramsey
configuration (see footnote \ref{fn:ram} and Fig.~\ref{f:ramsey}): in
fact, also the NOON states exhibit a special sensitivity with respect
to the transformation which encodes the random variable $\varphi$
(indeed NOON get transformed into output states $(e^{-i \varphi N/2}
|N,0\rangle+e^{i\varphi N/2}|0,N\rangle)/\sqrt{2}$ in which the phase
$\varphi$ result effectively multiplied by a factor $N$).  With this
choice Eq.~(\ref{QCRN33}) yields a lower bound of the form
 \begin{eqnarray}
  \delta \varphi_\nu \geqslant 
  \frac{1}{\sqrt{\nu} \; N}  \label{QCRN44} \;,
\end{eqnarray}
which, for any given $N$, is achievable in the limit of large $\nu$,
e.g. via maximum-likelihood estimation based on the photo-counting
statistics at the outport ports of the interferometer.  The bound
(\ref{QCRN44}) shows a $N^{-1/2}$ enhancement with respect to more
standard estimation approaches where, for instance, the input ports of
the interferometer are fed with coherent states of average photon
number $N$ (these procedures show a SQL scaling
$\delta \varphi_\nu = {1}/{\sqrt{\nu \;N}}$ in which, basically, {\em
  all} the $\nu N$ photons contribute independently in the estimation
process). For this reason Eq.~(\ref{QCRN44}) can be seen as the
quantum optical counterpart of the Heisenberg bound of
Eq.~(\ref{HEIS}).

The attainability of the bound~(\ref{QCRN44}) requires some extra
considerations.  First of all it assumes the ability of creating NOON
states. This is possible via some rather complicated optical
schemes~\cite{kok1,pryde,cable1,MITC04,LAMA01} which have only been
implemented in highly refined, but post-selected
experiments~\cite{milena,sara,MITC04,naga,OKA08,kacprowicz}.  However
states which possess a high fidelity with the NOON states, also for
large values of $N$, can be simply obtained by mixing a squeezed
vacuum and a coherent state at a beam
splitter~\cite{holgernoon,holger,itai}. Introducing such states at the
input of a Mach-Zehnder, a scaling $N^{-1}$ in the average photon
number of photons employed in a given experimental run can be achieved
\cite{cavesprd}. To do so however one needs to employ the proper
estimation process~\cite{PEZZE08}: a point which is not sufficiently
stressed in literature where often suboptimal performances can be
associated with a poor processing of the measurement outcomes.
Analogous $N^{-1}$ performances can also be achieved by employing
different sources and/or by using estimator functions which are
simpler than the maximum-likelihood approach.  Typically these schemes
are based on adaptive strategies where the parameter $\varphi$ is
pushed toward an optimal working point which guarantees higher
performances, e.g.~two recent proposals are Refs.~\cite{MONR06,cable} 
while a list of older ones are in  Refs.~\cite{GIOV04,rosetta}. 
Alternative schemes instead are
based on sequential strategies of Fig.~\ref{f:seq}
 where a {\em single} photon pulse
probes recursively the phase shift $\varphi$ by passing multiple times
through the delay line~\cite{luis,GIOV06,DEBU05,HIGG07,JI08}.

The achievability of the bound~(\ref{QCRN44}) for finite values of
$\nu$ is also non
trivial~\cite{DURK07,HAYA10-1,PEZZE08,VBRAU92,VBRAU92-1,VHRAD96,PEZZE06,VHRAD05,PEZZE07,LANE93}.
Several numerical analysis supported the evidence that Heisenberg-like
$N^{-1}$ scaling should be achievable in the limit of large $N$ also
for $\nu=1$, where Eq.~(\ref{QCRN44}) would yield a scaling analogous
to the strong Heisenberg bound (see the previous section).  In
particular Refs.~\cite{PEZZE08,PEZZE06,PEZZE07,VHRAD05} studied the
asymptotic behavior of the confidence of the error probability by
adopting a Bayesian estimation strategy~\cite{VHRAD96}.  A recent work
by Hayashi~\cite{HAYA10-1} however appears to settle the problem by
showing that while for $\nu=1$ the bound of Eq.~(\ref{QCRN44}) is not
exactly achievable, one could still reach an asymptotic $N^{-1}$
scaling for the RMSE by adopting a min-max optimization
approach~\cite{HAYA10}.

\section*{Filtering protocols}
Instead of going through the trouble of creating the complex and
fragile quantum states necessary for the quantum enhancements of
quantum metrology, some protocols have been proposed that use easy to
create and robust classical states, and then filter (post-select) the
high-resolution states at the measurement stage. The basic idea is to
employ retrodiction~\cite{pregnelpegg,resch}: once a high-resolution
quantum state has been detected at the output, one can interpret the
whole experiment as having employed such a state since the input. This
is a consequence of the fact that the wave-function collapse can be
placed at an arbitrary time between probe preparation and
measurement~\cite{pregnelpegg,peggbarnettjeffers}.

The filtering intrinsic in such protocols implies that part of the
resources available at the onset are wasted: the system is sampled
with many more resources than those actually employed for the
parameter estimation.  Moreover, since the system is sampled with
classical states that present no quantum correlations, it is 
clear from the preceding sections that the SQL
cannot be beaten: there is no increase in resolution over the optimal
{\em classical} strategy that could employ {\em all} the resources
(without filtering). For this reason, filtering protocols cannot be
considered proper quantum metrology protocols, according to the
definition we have given above.  Nonetheless they can be extremely
useful in the common case when the optimal classical strategy that
employs all the resources that have sampled the system is impractical.
Moreover, in practical situations efficiency considerations rarely
play a role, whereas robustness to noise is paramount.  Classical
states are, by definition, the most robust ones. In addition, there
are situations where filtering methods achieve tasks that would be
impossible with purely classical strategies, and post-selecting on the
high-resolution quantum states is often very simple.

In Ref.~\cite{resch} the theory of filtering protocols for phase estimation
is developed, and necessary conditions are given to distinguish
super-sensitivity (i.e.~the error in the estimation is lower than
allowed using classical resources) from super-resolution (i.e.~the
error in the estimation is lower than what would be allowed by a
classical procedure that uses only the resources that the filtering
retains).  Filtering protocols can achieve super-resolution (with an
appropriate measuring strategy) but cannot achieve the
super-sensitivity of quantum metrology proper.

\section*{Beyond the Heisenberg bound: nonlinear estimation strategies }

Several
Authors~\cite{LUIS04,BELT05,LUIS07,BOXI07,BOXI08,BOXI08-2,WOOL08,ROY08,CHAS09,CHOI08,MALD09,REY07,TILM10,RIVA08,RIVA10,NAPO10,SHAB10}
have recently considered the possibility of using nonlinear effects to
go beyond the $N^{-1}$ Heisenberg-like scalings in phase estimation
problems. These new regimes have been called ``super-Heisenberg''
scalings in Ref.~\cite{BOXI08-2}, but a proper accounting of the
resources shows that they are fully compatible with the analysis
presented in the previous sections, see e.g. Ref.~\cite{kok}.
Ultimately the idea of these proposals is to consider settings where
the unitary transformation that ``writes'' the unknown parameter $x$
into the probing signals, is characterized by many-body Hamiltonian
generators which are no longer extensive functions of the number of
probes employed in the
estimation~\cite{BOXI08-2,BOXI07,BOXI08,CHAS09,NAPO10,ROY08,TILM10,SHAB10,REY07,CHOI08}
or, for the optical implementations which yielded the
inequality~(\ref{QCRN33}), in the photon number operator of the input
signals~\cite{LUIS04,BELT05,LUIS07,RIVA10,RIVA08,WOOL08,MALD09,ROY08}.
Consequently, in these setups, the mapping $(e^{- i x H})^{\otimes n}$
which acts on the input states $\rho_0^{(n)}$, gets replaced by a
transformations of the form $e^{-i x H^{(n)}}$ which couples  the
probes non
trivially. Accordingly the minimization of
Eq.~(\ref{QCRN33}) is no longer forced to obey to the
inequality~(\ref{QCRN44}).  For instance, RMSE with scalings of the
order $\propto N^{-k}$ can be obtained when using Hamiltonians that
involves $k$-system interactions between the probes~\cite{BOXI07},
while $\propto 2^{-N}$ scaling can be achieved by introducing an
exponentially large number of coupling terms~\cite{ROY08}.  Proposed
implementations include scattering in Bose
condensates~\cite{BOXI08-2,CHOI08}, Duffing nonlinearity in
nano-mechanical resonators~\cite{WOOL08}, two-pass effective
nonlinearity with an atomic ensemble~\cite{CHAS09}, Kerr-like
nonlinearities~\cite{LUIS04,BELT05,LUIS07,RIVA10,MALD09}, and
nonlinear quantum atom-light interfaces~\cite{NAPO10}.

\section*{Quantum metrology with noise}

The study of noisy quantum metrology is a special case of parameter
estimation for channels, where the map $\Phi_x$ which describes the encoding of the unknown parameter $x$,
contains also the description of the noise tampering with the process. Hence, even though 
 very few general results are known, many results
detailed in the previous sections  can be used to characterize noise effects. 
 Non trivial examples are known  where a Heisenberg scaling can be
retained even in the presence of noise~\cite{JI08}. On the other hand 
a simple application of the programmable criterion  of  Ref.~\cite{JI08}  shows that even a small amount
of depolarizing noise is sufficient to ruin any sub-shot-noise performances one could reach by entangling the probing systems.
Typical quantum metrology protocols are indeed extremely sensitive to noise.
For instance, the incoherent loss of a single photon in a NOON state
transforms it into a statistical mixture 
$(|N-1,0\rangle\langle N-1,0|+|0,N-1\rangle\langle 0,N-1|)/2$ 
which is useless for
phase-sensing. Such 
extreme sensitivity to losses implies that this
state cannot be used in any practical situation~\cite{gilbert,kaus,holger}. Is it then really possible to outperform
classical strategies in practical phase sensing~\cite{bananat}?

Surprisingly, it has been shown that asymptotically it is possible to
do so only by a constant factor~\cite{dobrza,durkin}: for 
any nonzero loss, for sufficiently high number of photons $N$ the
scaling of the optimal phase sensing is proportional to the scaling of
the shot noise $\propto N^{-1/2}$.
While this means that quantum approaches are useful in highly controlled
environments~\cite{durkin} (such as for gravitational wave
detection~\cite{cavesprd}), they only allow for very small enhancements
in free-space target acquisition~\cite{durkin}. Nonetheless, the shot
noise can be beaten~\cite{mandm} and the optimal states to do so in the
presence of loss have been calculated numerically using various
optimization techniques for fixed number of input
photons~\cite{banasz,banasz1} and for photon-number
detection~\cite{optdowlin}. 
A post-selected proof of principle experiment that
employs some of these optimal states was recently
performed~\cite{kacprowicz}. Note that, for very low values of loss,
NOON states retain their optimality~\cite{banasz,optdowlin}, and can
be approximated by states that are easy to
generate~\cite{holger,holgernoon}. Also, a very simple proposal based
on parametric downconversion which can be realized without
post-selection was proposed in~\cite{cable}: it can achieve the
Heisenberg bound for low loss and degrades gracefully with noise. In
the case in which there is a large amount of loss {\em after} the
sample has interacted with the light probe, even achieving the shot
noise limit might become cumbersome (most of the photons that
interacted with the phase shifter and contain phase information are
lost). A simple strategy that amplifies the signal before detection
and can asymptotically achieve the shot noise was experimentally
tested in~\cite{fabio}. In contrast, the optimal states for the
sequential (or multi-round) interferometry~\cite{luis,HIGG07} have not
been obtained so far in the presence of noise~\cite{bananat}. However,
as an alternative to the fragile two-mode states, some more robust
single mode states were also analyzed, e.g.
pure Gaussian states in the presence of phase
diffusion~\cite{parisphasediffusion}, mixed Gaussian states in the
presence of loss~\cite{tapia}, or single mode variants of the two mode
states~\cite{cillis}. In contrast to the two mode case (where the
phase is the relative one between the two modes) here the phase is
measured relative to a strong classical signal (using heterodyne or
homodyne measurements) or similar strategies.

Historically, the first analysis of a quantum metrology
protocol~\cite{wineland} in the presence of noise was performed
in~\cite{chiara}, where frequency measurements are analyzed in the
presence of dephasing. Interestingly, a phase transition was shown:
for any nonzero value of dephasing, the maximally entangled state
suddenly ceases to present any advantage over a classically correlated
state. (An analogous result for magnetometry exists~\cite{yashc}.) We
briefly present this result, as it is instructive of the subtleties
encountered by quantum metrology when noise sources are considered.

Consider the Ramsey setup of Fig.~\ref{f:ramsey} when a dephasing
process acts on the atomic probes.    
For the conventional setup the probability of finding the probe in the
initial state after the application of the phase shift becomes
$p=(1+e^{-\gamma t}\cos\varphi)/2$, where $\gamma\geqslant 0$ measures
the dephasing rate and $t$ is the time elapsed from the state
preparation up to the measurement readout. Analogously, for the
quantum setting we get $p_{ent}=(1+e^{-\gamma N t}\cos N\varphi)/2$,
where the factor $N$ in the exponent derives from the exponentially
greater sensitivity of the entangled state to the dephasing. The
corresponding RMSE (using the equations derived in
Fig.~\ref{f:ramsey}) is $\delta\varphi_n=(\frac{e^{2\gamma
    t}-\cos^2\varphi}{n\sin^2\varphi})^{1/2}$, and
$\delta\varphi_{n}=(\frac{e^{2\gamma N
    t}-\cos^2N\varphi}{\nu N^2\sin^2N\varphi})^{1/2}$, respectively.  For
$\gamma>0$ they are both minimized for $t=0$ (namely, we have to use a
measurement procedure which is as fast as possible), and one recovers
the typical ${N}^{1/2}$ enhancement of quantum metrology also in the
presence of dephasing. However, if one wants to measure a
frequency~\cite{wineland,chiara} $\omega$, then the phase factor is
$\varphi=\omega t$. In this case, the errors of the separable and
entangled procedure are respectively $\delta\omega_n=
(\frac{e^{2\gamma t}-\cos^2\omega t}{nt^2\sin^2\omega t})^{1/2}$ and
$\delta\omega_{n} = (\frac{e^{2\gamma N t}-\cos^2N\omega
  t}{\nu N^2t^2\sin^2N\omega t})^{1/2}$.  For $\gamma=0$, we have
$\delta_n\omega=1/t\sqrt{n}$ and $\delta\omega_{n}=1/(tN\sqrt{\nu})$. The
optimization over $t$ says that our measurement must be as long as
possible. Choosing the same $t$ in the two cases, the entangled
protocol is $\sqrt{N}$ more precise.  Interestingly, for $\gamma>0$
the two situations cannot be anymore optimized on the same $t$. For
example, if the frequency is $\omega=\pi\gamma$, it is easy to see
that one has to choose ${t}=(2\gamma)^{-1}$ to optimize the separable
procedure and ${t}=(2N\gamma)^{-1}$ to optimize the entangled one.
With these choices, one obtains for the two cases
$\delta\omega_n=\sqrt{e(2\gamma)^2/n}$ and
$\delta\omega_{n}=\sqrt{e(2\gamma)^2/\nu}$. We must assign equal resources
to the two protocols, so we can repeat times the entangled procedure
$N$ times as it is $N$ times faster. This recovers a factor
$1/\sqrt{N}$ also for the entangled procedure, whose error then
matches the separable one, $\delta\omega_n$.  In~\cite{chiara,kitag} it
is shown that using non-maximally entangled states one can beat the
separable procedure by $\sim 40\%$, which was proven to be the ultimate gain. 

Summarizing, the estimation of the phase $\varphi$ is unaffected by
the presence of dephasing noise (as long as the measurement time $t$
can be chosen short enough): the full ${N}^{-1/2}$ enhancement of
quantum metrology is retained.  On the contrary, the estimation of the
frequency $\omega=\varphi/t$ is strongly affected by the same noise:
basically all quantum enhancement is lost for any value of $\gamma$.
This is connected to the fragility of the entangled resource in the
presence of noise: the entangled procedure must be performed $N$ times
faster, which reduces the precision in the estimation of $\omega$ by a
factor $N$, thus achieving the same final yield as the separable
protocol. Note, however, that if the observation time ${t}$ in atomic
clocks is constrained by experimental issues (typically, by the local
oscillator fluctuations), then entanglement allows a sub-shot noise
scaling~\cite{nist} which scales as $N^{-2/3}$~\cite{lukin}. A
generalization of the frequency measurement theory was given
in~\cite{shaji}, which covers many noisy estimation measurements when
the measurement duration time and production rate of the probes are
considered as relevant resources.

VG acknowledges support from the FIRB-IDEAS project under the contract
RBID08B3FM and support of Institut Mittag-Leffler (Stockholm), where
he was visiting while part of this work was done. SL acknowledges support from 
DARPA, NEC, NSF, and ENI via the MIT Energy Initiative.

 \end{document}